# Band gap unification of partially Si-substituted single wall carbon nanotubes

Pavel V. Avramov[1, 2*], Pavel B. Sorokin[2, 3], Alexander S. Fedorov[2],

Dmitri G. Fedorov[4] and Yoshihito Maeda[1, 5]

[1] Takasaki-branch, Advanced Science Research Center, Japan Atomic Energy Agency, Takasaki, 370-1292, Japan

[2] L.V. Kirensky Institute of Physics SB RAS, 660036 Krasnoyarsk, Russian Federation

[3] Institute of Biochemical Physics of RAS, 119991 Moscow, Russian Federation

[4] Research Institute for Computational Science, National Institute of Advanced Industrial Science and Technology (AIST), Tsukuba, 305-8568, Japan

[5] Department of Energy Science and Technology, Kyoto University, Sakyo-ku, Kyoto 606-8501, Japan

The atomic and electronic structure of a set of pristine single wall SiC nanotubes as well as Si-substituted carbon nanotubes and a SiC sheet was studied by the LDA plane wave band structure calculations. Consecutive substitution of carbon atoms by Si leads to a gap opening in the energetic spectrum of the metallic (8,8) SWCNT with approximately quadratic dependence of the band gap upon the Si concentration. The same substitution for the semiconductor (10,0) SWCNT results in a band gap minimum (0.27 eV) at ~25% of Si concentration. In the Si concentration region of 12–18%, both types of nanotubes have less than 0.5 eV direct band gaps at the Γ-Γ point. The calculation of the chiral (8,2) SWSi$_{0.15}$C$_{0.85}$NT system gives a similar (0.6 eV) direct band gap. The regular distribution of Si atoms in the atomic lattice is by ~0.1 eV/atom energetically preferable in comparison with a random distribution. Time dependent DFT calculations showed that the silicon substitution sufficiently increases (roughly by one order of magnitude) the total probability of optical transitions in the near infrared region, which is caused by the opening of the direct band gap in metallic SWCNTs, the unification of the nature and energy of band gaps of all SWCNT species, the large values of $\langle Si3p|r|Si3s \rangle$ radial integrals and participation of Si3$d$ states in chemical bonding in both valence and conductance bands.

*Correspondent author,

e.mail: avramov.pavel@jaea.go.jp

phone: +81 27 346 9670

fax:     +81 27 347 2501

# I. INTRODUCTION

The most promising way to modify the electronic structure of single wall carbon nanotubes (SWCNTs) is a modification of the carbon lattice by some physical or chemical defects. [1-12] For example, a scanning tunneling spectroscopy (STS) study [1-3] of the electronic structure of metallic (13,13) SWCNT with defects demonstrated a number of additional spectroscopic features in the density of electron states (DOS). The chemical agents [4,5] like $O_2$ and $NO_2$ increased the DOS at the Fermi level and converted semiconducting nanotubes into metallic ones. Based on the *ab initio* calculations [6] of the SWCNT/$O_2$ system it was shown that extra charge carriers (holes) appear at the Fermi level closing the bandgap in semiconductor *p*-type nanotubes.

Since the discovery of the band gap photoluminescence (PL) in the 800–1600 *nm* wavelength range, [13] the semiconducting single-walled carbon nanotubes (SWCNTs) emerged as systems with a great potential for nanotechnological applications. The low photoluminescence quantum yield [13,14] is one of the major obstacles towards a widespread use of the nanotubes in optical devices. The PL study [15] shows the strong dependence of the PL quantum yield on the presence of defects in the carbon SWCNT lattice.

Using *ab initio* DFT approach [9] it was shown that the lattice and chemical defects lead to a sufficient change in the electronic structure of the semiconducting SWCNT closing the band gap in the DOS. The chemical functionalization like creation of C-H bonds outside and inside armchair and zigzag nanotubes [10,12,16] leads to a sufficient widening of the SWCNT's band gap. In contrast, the nitrogen impurities [11] result in the formation of the embedded electronic states at the Fermi level.

Some attempts were made to synthesize the semiconducting nanotubes (and other nanostructures like nanorods, nanofibers, nanocones and nanowires) on the silicon carbide (SiC) base. [17-31] The atomic structure of the resulting silicon carbide single wall nanotubes (SWSiCNT) remains very close to the original shape of the initial SWCNTs. The first theoretical investigation of the SWSiCNTs with different Si/C ratio was done using cluster DFT [32] and PBC GTBMD approaches. [33] It was shown that SWSiCNTs have two stable atomic structure types close in energy but significantly different in electronic properties. The more energetically stable Type-1 consists of altering Si and C atoms and Type-2 possesses the $Si_2$-$C_2$ alteration. [33] Using the PBC DFT approach [34] it was shown that all SWSiCNTs are semiconductors with the band gap varying from 0.2 (zigzag structures with narrow diameters) to 2.0 eV (large armchair structures). All zigzag tubes have a direct band gap at the Γ point of the Brillouin zone (like zigzag and semiconducting chiral SWCNTs [35-37]), whereas armchair and chiral ones have an indirect band gap. In case of hydrogen decoration of the C-sites, all SWSiCNTs display the *n*-type semiconducting properties in contrast with the Si-site ones (*p*-type).

The opposite behavior of the deep hydrogenation of Si- and C-sites of the armchair and zigzag SWSiCNTs was reported.[38] Full hydrogen coverage of Si atoms of the (8,0) and (6,6) structures leads to metallic states, whereas the C-type hydrogenation leads to an increase of the band gap up to 2.6 eV for the (8,0) and a decrease up to 1.47 eV for (6,6). The hydrogenation of all atoms increases the band gap of all species. The electronic structure of $XH_3$–radical decorated SWSiCNTs (X=C, Si) was studied by LDA PBC calculations.[39] It was shown that the modification of the electronic structure highly depends on the adsorption sites rather than the $XH_3$ species (semiconducting *n*-type for the C-site absorption and *p*-type for Si-site).

The Si- doped SWCNTs as well as bulk silicon were studied by LDA PBC calcuations.[40-42] It was shown that the carbon-induced states are highly localized and form a deep level in the forbidden band gap. The atomic and electronic structure of $Si_{1-y}C_y$ ($y<0.5$) was calculated using LDA FP-LAPW method.[41] At a small concentration of carbon in crystal lattice the SiC alloys are semiconductors with very small direct band gap (silicon itself has indirect band gap). Such direct-indirect band gap transformation for other systems (InP, $Ga_xIn_{1-x}As$/InP, GaAs/AlAs) was detected experimentally.[43-45]

Summarizing the works mentioned above, at present there is no way to produce 1D semiconducting nanostructures like nanotubes with small uniform direct band gaps. The pristine SWCNTs demonstrate wide variations of the band gaps from 0 up to 1 eV and more [2] with direct and indirect band gaps. The chemical functionalization of both SWCNTs and SWSiCNTs leads to the wide band gap electronic structure, and the selective functionalization produces both metallic and semiconducting states depending upon its chirality and type. In contrast to pure silicon, a small carbon substitution in the bulk silicon makes the band gap direct and leads to a sufficient decrease of its width.

The detailed study of the electronic structure and finding the possible ways of the unification of the SWSiCNT band gap for various nanotechnological applications are the main goals of the present paper. Unification here denotes the means of producing the nanotubes with the band gap nearly independent of the structure and determined almost exclusively by the Si concentration.

II. METHOD OF CALCULATIONS AND OBJECTS UNDER STUDY

We studied the atomic and electronic structure of a number of partially Si-substituted SWCNTs with different Si content: (a) 5.2, 16.7, 25.0 and 43.8% for (8,8) nanotube (NT); (b) 15% for (8,2) NT; (c) 1.25, 12.5, 25.0, 35.0 and 37.5% for (10,0) NT; (d) a set of pristine armchair ((4,4), (5,5), (6,6), (8,8) and (15,15)), zigzag ((10,0)), chiral ((6,4) and (8,2)) Type-1 SWSiCNTs and a correspondent plane Type-1 SiC sheet.

For all calculations of the periodic systems the Vienna *ab-initio* simulation package (VASP) [46, 47] was used. The total energy code is based on LDA approximation [48, 49] with the Ceperley-Alder exchange-correlation functional, [50] plane-wave basis sets and ultrasoft pseudopotentials (PP). [51] The pseudopotentials allow one to significantly reduce the maximal kinetic energy cutoff without loss of accuracy. All electronic structure calculations were performed within the PBC approximation (LDA PW PP PBC). The geometry optimization was carried out until the forces acting on all atoms become lower than 0.05 eV/Å. To produce the total density of states (DOS), a Gaussian broadening for the electronic occupation numbers was used with a smearing width of 0.1 eV.

As the unit cell of (6,4) SWSiCNT contained a large number of atoms, the electronic structure was calculated by crimped strip structure method. [52, 53] The basic idea of this approach is the partial unrolling of quasi one-dimensional nanotube structure into a two-dimensional crimped strip structure to reduce the unit cell dimensions. Additionally, the number of atoms in the new unit cell is reduced by some integer factor. These steps lead to a very considerable reduction of computational cost for nanotubes with large diameters or chiral structures.

To calculate partially Si-substituted (8,8) SWCNT we used the 96 atom super cell containing five (5.2%) and 16 silicon atoms (16.7%) and the 64 atom super cell containing 16 and 30 Si atoms (25.0% and 43.8% respectively). The unit cells of pristine (8,8) SWCNT and Type-1 SWSiCNT (50% concentration) consisted of 32 and 64 atoms correspondingly.

The same procedure was developed to elaborate the slab models of the partially Si-substituted (10,0) nanotube. The unit cells consisted of 1 Si and 79 C atoms (1.25%), 10 Si and 70 C (12.5%), 20 Si and 60 C (25%), 28 Si and 52 C (35.0%), 30 Si and 50 C (37.5%). The unit cells of the pristine (10,0) SWCNT and Type-1 (10,0) SWSiCNT (50% concentration) contained 40 atoms. To study the role of the regular Si distribution, a slab model (16 Si and 48 C atoms, concentration 25%) with quasirandom Si atom locations was developed. The unit cell of 15% Si substituted (8,2) NT consisted of 9 silicon and 47 carbon atoms.

In order to study the quantum efficiency of the proposed nanotubes as elements of optical devices, we used the time dependent [54] (TD) DFT (PBE0/6-31+G**) [55-57] approach within the adiabatic approximation and GAUSSIAN03 code [58] to evaluate an enhancement of the total dipole transition intensities between the valence and conductance bands of the Si-substituted objects using cluster models ($C_{54}H_{18}$ and $Si_6C_{48}H_{18}$, fragments of pristine and 11% silicon substituted (9,0) SWCNT; hydrogen atoms were added to the C-C dangling bonds). The atomic structures of both clusters were optimized using PBE0 hybrid potential, 6-31+G** basis set and analytic energy gradients. Previously, it was shown[59] that for the semiconducting 1D SWCNTs (and both clusters definitely have zero DOS at Fermi level because of the finite atomic structure and terminal hydrogen atoms as well) the hybrid PBE0 potential [55, 56, 59] gives excellent results in determination of the equilibrium atomic structure, the width of the forbidden gap and density of electronic states near the Fermi level. It is necessary to note that the hybrid PBE0 potential gives close results to the fourth-order Møller-Plesset perturbation theory (MP4) due to the admixture of 25% of the exact Hartree-Fock exchange. [56] Later we will discuss shortly the importance of hybrid potentials for the TD DFT calculations of optical transitions. [54]

In the works of Refs. 54, 60-65 it was shown that even at the LDA level (Gaussian type based PBC LDF [65] and PW LDA [63, 64]) it is possible to describe the main features of the optical spectra of the SWCNTs. However, based on the Green's function (GF) approach, the importance of the exciton interactions in the mechanism of the optical spectra formation of SWCNTs was demonstrated. [66]

At present, two main approaches [67] beyond the one electron approximation (TD DFT and GF) are used to calculate the optical spectra of relatively large systems taking into account the local field effects and correct exchange-correlation potential. The first one (TD DFT) is preferable because of its high accuracy and reasonable computational cost. [54, 66-68]

The TD DFT approach gives good results for the optical spectra even in the PW or Gaussian-type based LDA approximation. [60, 68] The introduction of hybrid potentials like B3LYP [54] can sufficiently improve the results even at the 3-21G level of theory for the $C_{70}$ molecule. Nevertheless, using better basis sets like 6-31G*, cc-pVTZ [62], 6-31+G* or DZP [61] can consequently improve the results. Comparing the 6-31G* and cc-pVTZ results, as well as 6-31+G* and DZP, one can conclude that even a modest basis set (6-31G*) can give good results for the adiabatic electron affinity and the first ionization potential of $C_{60}$ molecule. [62]

Finally, to evaluate the role of pure atomic factors such as the values of radial dipole matrix elements ($\langle Si3p|r|Si3s \rangle$ and $\langle C2p|r|C2s \rangle$) in determination of the optical properties of the $SWSi_xC_{(1-x)}NTs$, we calculated the dipole transitions for Si and C atoms using the same TD DFT approach.

### III. RESULTS AND DISCUSSION

*a) Test systems*

To justify the quality of LDA PW PP PBC calculations of the 1D structures like partially silicon substituted carbon nanotubes, the atomic and electronic structure of some well known test systems like (8,8) and (10,0) SWCNTs, plane SiC sheet and a set of SWSiCNTs was calculated. The quality of the calculations is demonstrated by a comparison of the LDA PW PP PBC DOS for (8,8) SWCNT (Fig. 1) with theoretical Gaussian-type (G-type) PBC PBE 6-31G* results [59] and experimental STS spectrum. [2]

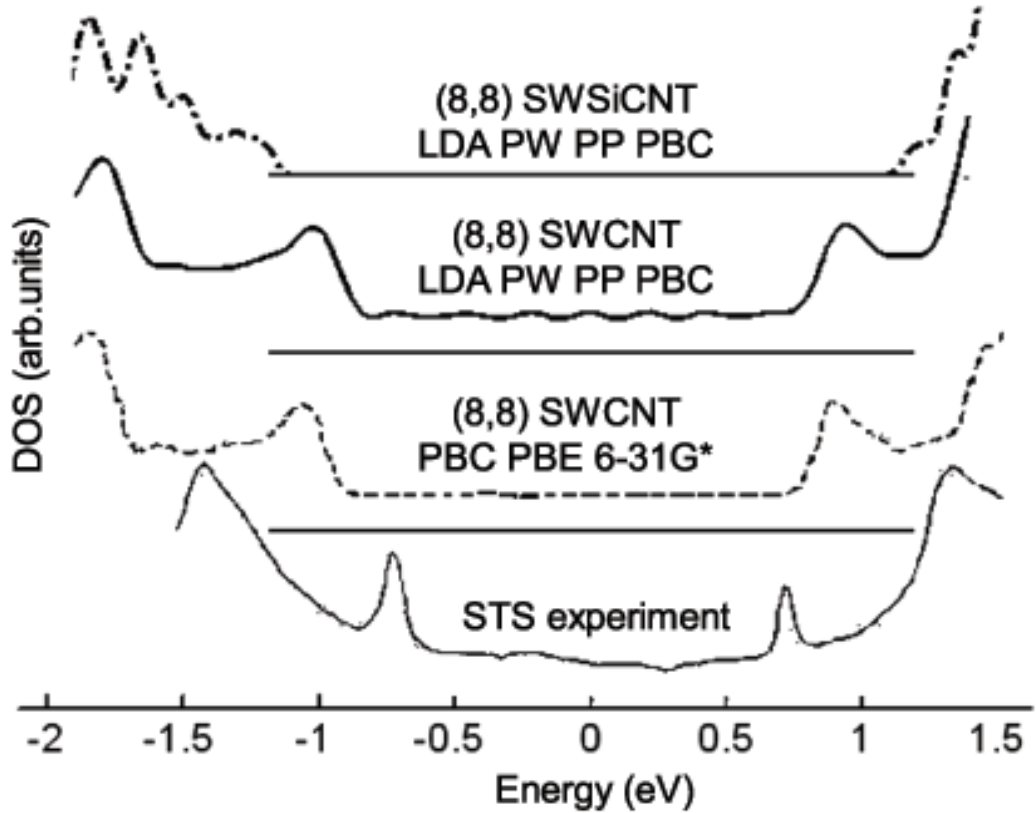

**Fig. 1. The theoretical LDA PW PP PBC, G-type PBE PBC [59] DOS and experimental [2] STS spectrum of pristine (8,8) SWCNT as well as the LDA PW PP PBC DOS of the pristine (8,8) SWSiCNT. The Fermi level energy is taken as zero for all tubes, except for SWSiCNT with $E_{Fermi}$ = -1.12 eV.**

For the metallic (8,8) SWCNT system both pure DFT Gaussian type PBE/6-31G* and LDA PW PP PBC methods give close values for the density of states and the energy difference between the first van-Hove singularities (VHS) (2.0 eV) with the same (0.5 eV) accuracy in describing the experimental density of states.[2]

The comparison of the LDA PW PP PBC and Gaussian type PBE/6-31G*[59] calculations of semiconducting (10,0) SWCNT shows that both methods give close values for the forbidden gap (0.75 and 0.8 eV correspondingly). In comparison with the experimental data,[2] the LDA PW PP PBC and Gaussian type PBE/6-31G* DOS (not presented in Fig. 1 for the (10,0) SWCNT, but can be found in Ref. 59 and below in the present work) have close shapes (relative energies and VHS intensities). Thus, based on the test calculations, it is expected that the quality of LDA PW PP PBC DOSes for a set of SWSiCNTs and partially silicon substituted SWCNTs should be quite reasonable.

The electronic band structure calculations of the Type-1 SiC plane sheet was performed using the LDA PW PP PBC approach. Obviously, the SiC sheet itself is a wide-gap semiconductor with the energy of the forbidden gap $E_{gap}$=2.6 eV. Assuming the small importance of the NT curvature in case of a quite large diameter, all pristine SWSiCNTs should be semiconductors.[69]

To confirm this result, we performed LDA PW PP PBC calculations of Type-1 pristine (or 50% Si concentration) armchair ((4,4), (5,5), (6,6) and (15,15)) SWSiCNTs (Fig. 2(a)) and some chiral ((6,4) and (8,2)) and zigzag (10,0) SWSiCNTs of a similar diameter (Fig. 2(b)). All species are semiconductors with the band gap from 1.7 ((8,2) SWSiCNT) up to 2.5 eV ((15,15) SWSiCNT). All SWSiCNTs retain the perfect shape of the atomic structures of the parent SWCNTs. As in the earlier work of Ref. 34, all chiral and armchair SWSiCNTs (Figs. 2a, 2b) have an indirect band gap, whereas the zigzag one (Fig. 2(b)) has a direct Γ-Γ band gap (the band structures are presented in the insets of Fig. 2). In comparison with GTBMD[33] calculations our LDA PW PP PBC calculations systematically underestimate (by more than on 1 eV) the band gap energies of SWSiCNTs. For the SiC structures it was shown[70-73] that for the LDA approach the systematic quasiparticle shift of the energy of forbidden gap is equal to 1.1 eV or more.

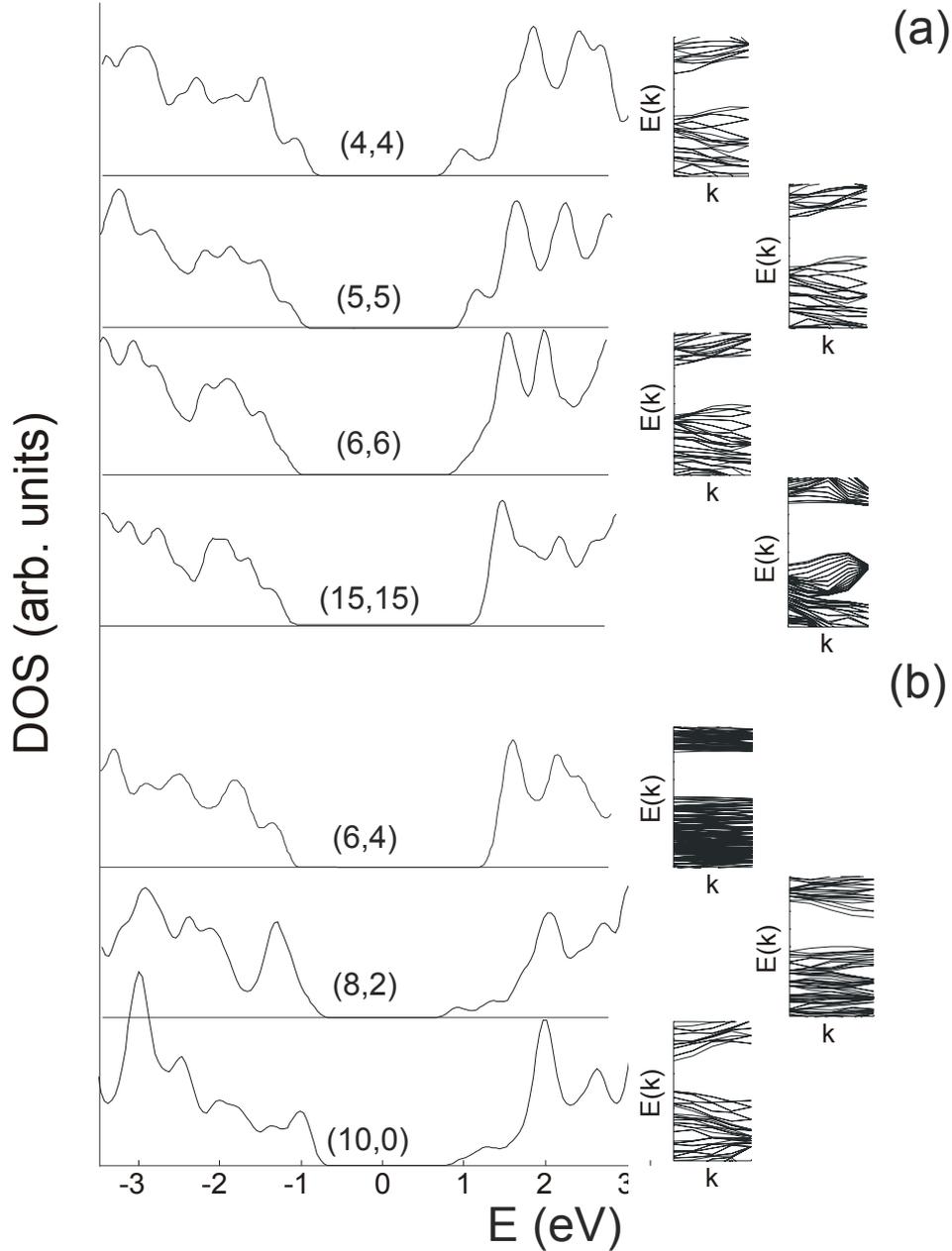

**Fig. 2. The theoretical LDA PW PP PBC DOS and the correspondent band structures for armchair (4,4), (5,5), (6,6) and (15,15) SWSiCNTs (a) and for chiral ((6,4) and (8,2)) and zigzag (10,0) SWSiCNTs of a similar diameter (b). The Fermi level energy is taken as zero.**

*b) Si- substituted SWCNTs*

To study the band gap change under the substitution of carbon atoms by Si we performed the electronic structure calculations of a number of Si-substituted carbon nanotubes of (8,8) armchair (5.2%, 16.7%, 25.0% and 43.8% of Si in the unit cells) and (10,0) zigzag SWCNTs (1.25%, 12.5%, 25.0%, 35.0% and 37.5% of Si in the unit cells) as well as pristine (8,8) and (10,0) SWCNTs and (8,8) and (10,0) SWSiCNT (Fig. 3, 4, 5). We selected the (8,8) and (10,0) based structures as examples to study the influence of silicon substitution because of the presence of the high quality experimental STS [2] and theoretical [59] PBC PBE0/6-31G* DOS of the (8,8) and (10,0) SWCNTs species.

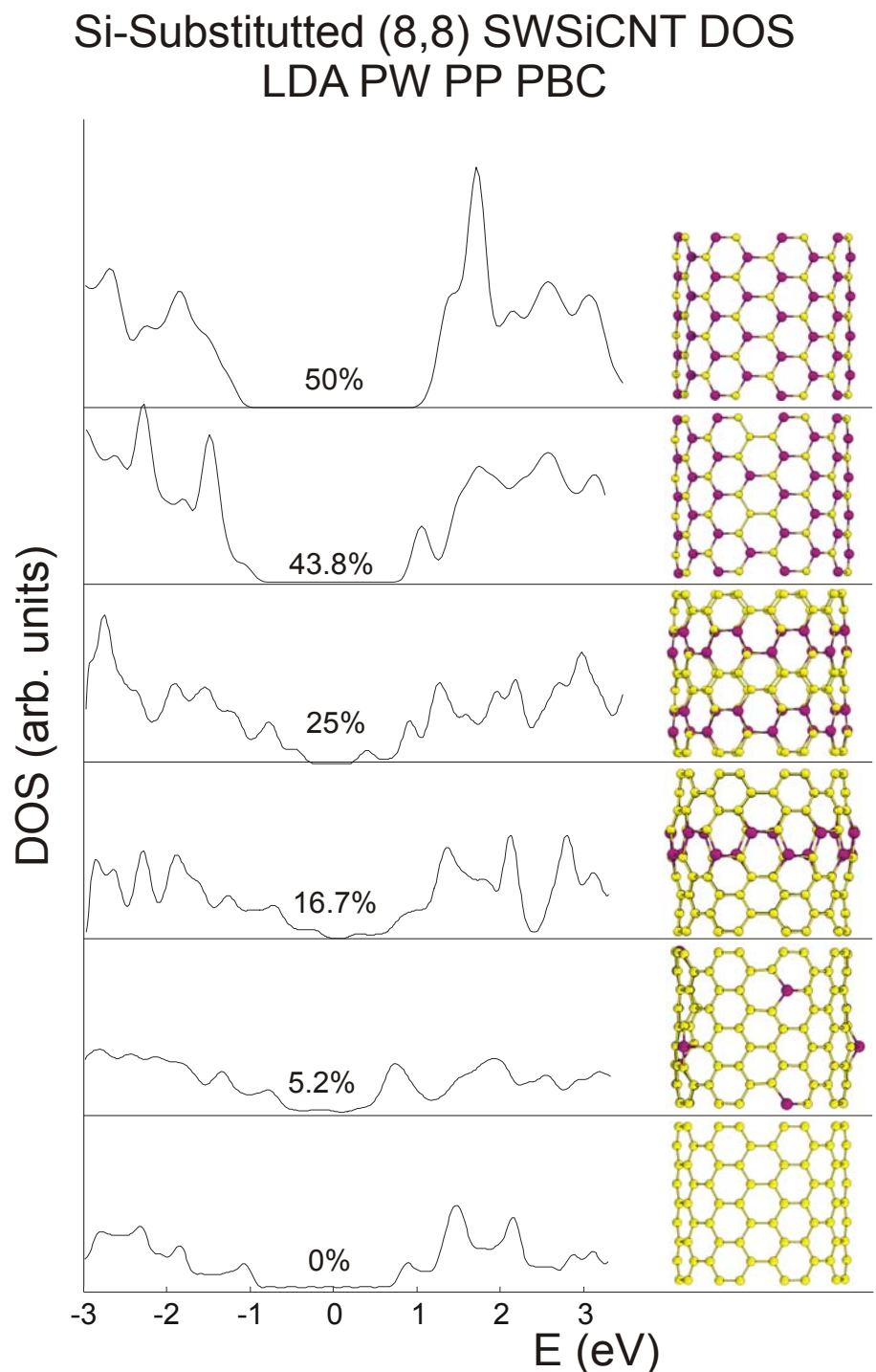

**Fig. 3. The theoretical LDA PW PP PBC DOS of a number of Si-substituted (8,8) armchair SWCNT (43.8%, 25.0%, 16.7% and 5.2% of Si in the unit cells) structures as well as pristine (8,8) SWCNT (0%) and (8,8) SWSiCNT (50%) (the Fermi level is taken as zero). The corresponding atomic lattices of the structures are shown on the right side of the figure (carbon atoms are in yellow (gray) and silicon atoms are in brown (black).**

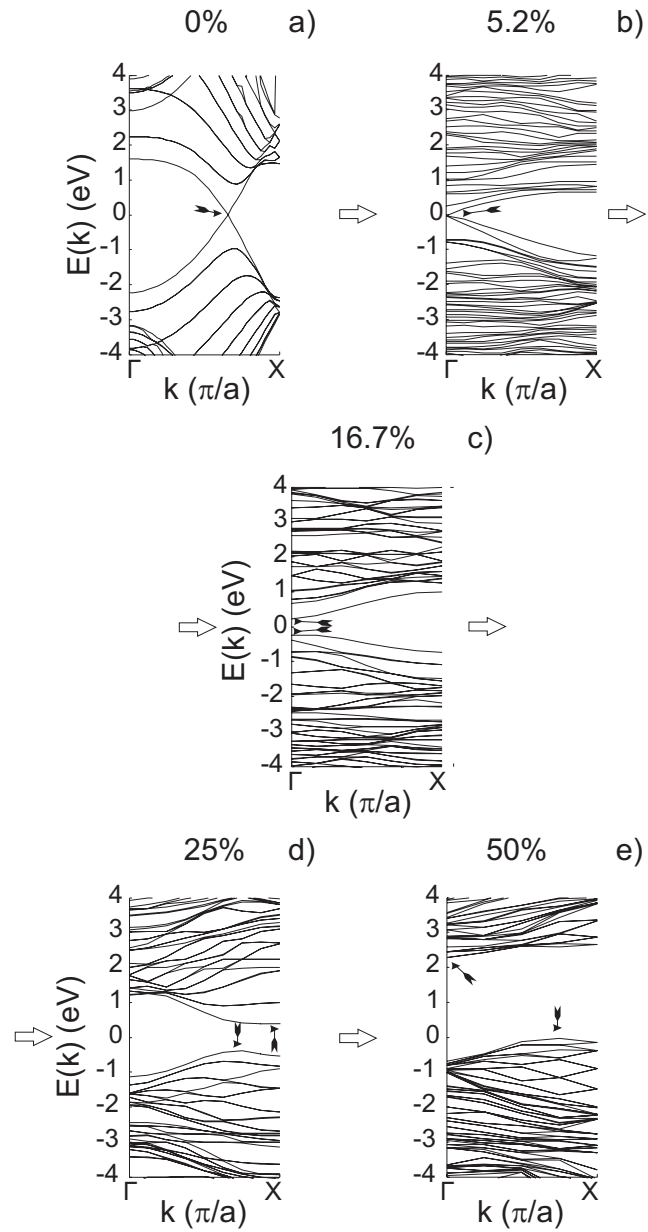

**Fig. 4.** The band structure of pristine (a) (8,8) SWCNT and (e) (8,8) SWSiCNT as well as 5.2% (b), 16.7% (c) and 25% (d) silicon substituted (8,8) SWCNT. The arrows show the highest/lowest points of valence/conductivity bands. It is clearly seen that the pristine (8,8) SWSiCNT (e) as well as 25% system (b) are indirect band gap semiconductors. The 16.7% system has a narrow direct band gap.

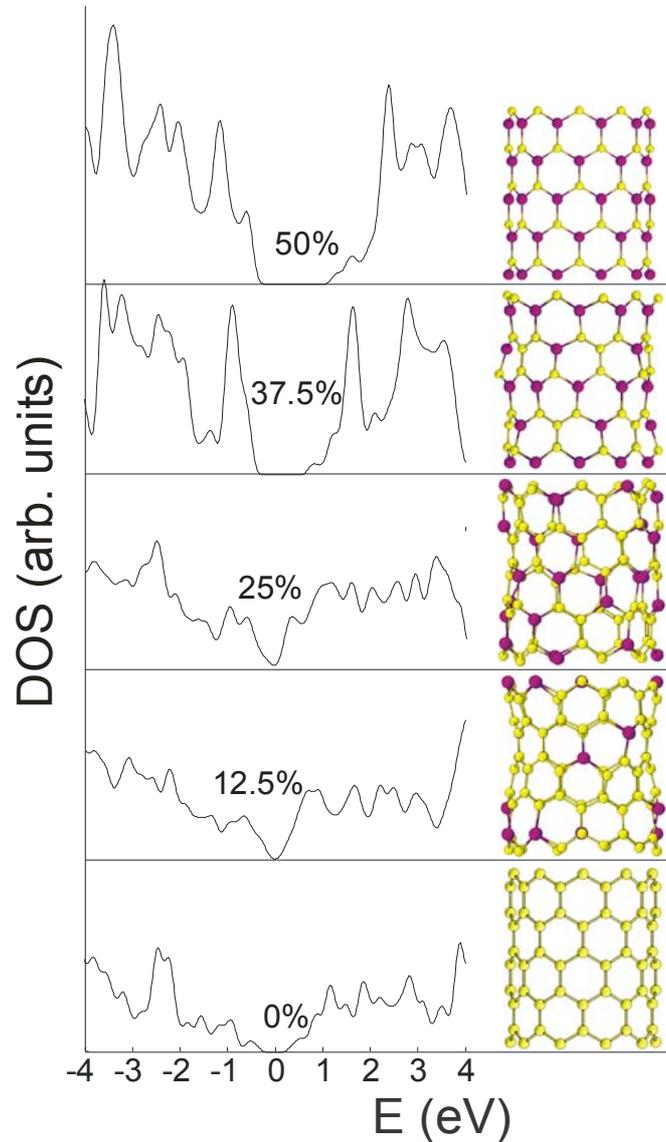

**Fig. 5.** The theoretical LDA PW PP PBC DOS of a number of Si-substituted (10,0) zigzag SWCNT (12.5%, 25.0% and 37.5% of Si in the unit cells) structures as well as pristine (10,0) SWCNT (0%) and (10,0) SWSiCNT (50%). The Fermi level energy is taken as zero. The corresponding atomic lattices of the structures are shown on the right side of the figure (carbon atoms are in yellow (gray) and silicon atoms are in brown (black).

The regular substitution does not affect the perfect atomic structure of the parent SWCNT systems in all cases. In Figs. 3 and 5, the corresponding regular substituted structures are presented on the right side. For these species the concentration dependence of the band gap is presented in Fig. 6.

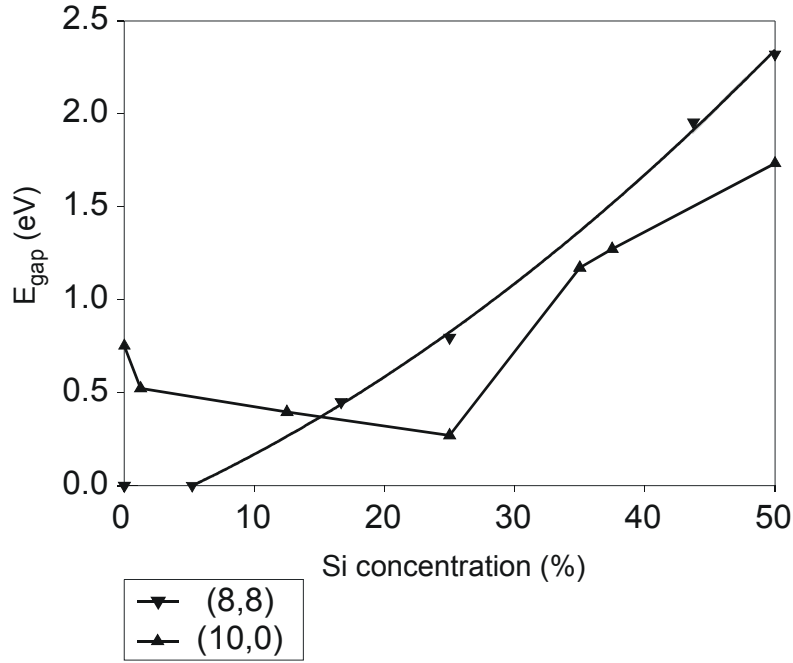

**Fig. 6. The concentration dependence of the band gap $E_{gap}$ of Si-substituted (8,8) (downward triangles) and (10,0) SWCNTs (upward triangles).**

Even relatively small concentration (5.2%) of silicon keeping the metallic nature of the (8,8) system (Fig. 3), leads to the energy shift (0.3 – 0.5 eV) of the first two VHS peaks in valence and conductivity bands followed by sufficient intensity changes (increasing the peak intensities at ~ -2 eV in the valence band and at 0.6 and 1.8 eV in the conductivity band and decreasing the peak intensity at 1.45 eV in the conductivity band). The DOS of 16.7% system demonstrates an appearance of new features at -0.15 eV and 0.25 eV in the valence and conductivity bands. These peaks reflect a presence of a sufficient amount of new Si-C chemical bonds formed with a participation of the Si$3d$-states. Other three VHS in the valence band in the energy range of -0.8 ÷ -2.0 eV keep the nature of the parent (8,8) SWCNT system. Keeping the VHS relative energies, the third peak intensity undergoes a sufficient change caused by the silicon contribution in the electronic structure. The perfect atomic structure of the species prevents the splitting of the peaks in the DOS. On the contrary, the shape of the conductivity band is totally changed because of the presence of the Si$3d$ contribution. Later we will show that Si$3d$-states play a really important role in the chemical bonding and dramatically affect the optical properties of the Si-substituted systems.

Because of the different nature of the chemical bonding, the pristine (8,8) SWSiCNT (50% concentration) DOS sufficiently differs from the pristine SWCNT (0% concentration) one (Fig. 3). The top of the valence band of the SWSiCNT can be characterized by the sufficient contribution of Si3*d*-states. The valence band DOS of the 43.8% system is close in the shape to the 50% one because of the high concentration of silicon. The energy positions and relative intensities of the peaks in the valence band DOS of the 25% system are sufficiently different from those with both low and high silicon content.

The conductivity band of the silicon substituted species strongly depends on the silicon concentration due to the presence of a large contribution of the Si3*d*-states and additional silicon core electrons, effecting the increase of the spatial extension of the silicon *s* and *p* orbitals. Because of the strong hybridization nature of C-C and Si-C bonds, even under small concentrations of silicon, the silicon-related states in both valence and conductivity regions can not be attributed to the embedded bands in the forbidden gap. This conclusion can be confirmed by the analysis of the band structure dispersion law (Fig. 4) of the species. Even a small (5.2%) concentration (Fig. 4b) of silicon leads to the fundamental reconstruction of the band structure of the (8,8) SWCNT (Fig. 4a) like preventing the band crossing and changing the nature of the vacant and valence bands in the whole energy region. The consequent increase of the silicon concentration leads to the sufficient rearrangements of the dispersion law of the systems. Because of all of these facts no linear dependence of the DOS upon the silicon concentration can be expected nor is observed.

The substitution of 5.2% of Si keeps the metallic nature of (8,8) SWCNT. The "metal-semiconductor" phase transition occurs after the 5.2% of the silicon concentration. From 5.2% and up to 50% of Si, a roughly quadratic dependence "Si concentration *vs* width of the band gap" is observed (Figs. 3, 6). The silicon impurities (or substitution centers) destroy the metallic nature of the (8,8) system smoothly converting it from a metal and a narrow band gap semiconductor into a wide band gap semiconductor.

The consequent Si-substitution of carbon atoms in the atomic lattice of the (10,0) SWCNT (Figs. 5, 6) leads to a decrease in the band gap from 0.75 eV (pristine SWCNT or 0% of Si) up to 0.27 eV (25% of silicon). The following Si substitution leads to an increase of the band gap up to 1.7 eV (50% of Si). In the 12 – 18% range of concentration (Figs. 3, 5, 6) the band gaps of both systems become close to each other (~0.4 eV) with direct band gap at the Γ point (Fig. 4). Even a small level of Si-substitution (5.2%) in (8,8) SWCNT shifts the band crossing from $\frac{1}{\sqrt{3}}\pi a$ to the Γ-point. Increasing the Si concentration leads to the direct band gap semiconducting structure at the Γ-point of the Brillouin zone and, at 25%, 43.8% (not presented on the figure) and 50% to the indirect band gap semiconducting structure. The Si substitution keeps the direct band gap at the Γ-point semiconducting structure of (10,0) SWCNT/SWSiCNT in the whole range of concentrations (0-50%).

The electronic structure calculations of the unit cell with random distribution of the silicon atoms (25%) do not show a significant change in the width of the bandgap for the correspondent perfect Si-substituted (8,8) SWCNT (1.04 eV for random structure and 0.80 eV for the regular one). The random distribution of silicon atoms in the atomic lattice leads to an elliptic cross-section of the unit cell and an increase in the energy of the system up to the 0.1 eV per atom.

To confirm the energy band gap concentration dependence (Fig. 6) for the case of chiral nanotubes we performed as an example the calculation of (8,2) SWSi$_{0.15}$C$_{0.85}$NT system (Fig. 7). The 15% silicon substitution leads to the direct band gap opening at $\frac{1}{\sqrt{3}}\pi a$ with the 0.6 eV width. This result is close to the data (Fig. 6) for the armchair and zigzag silicon substituted systems.

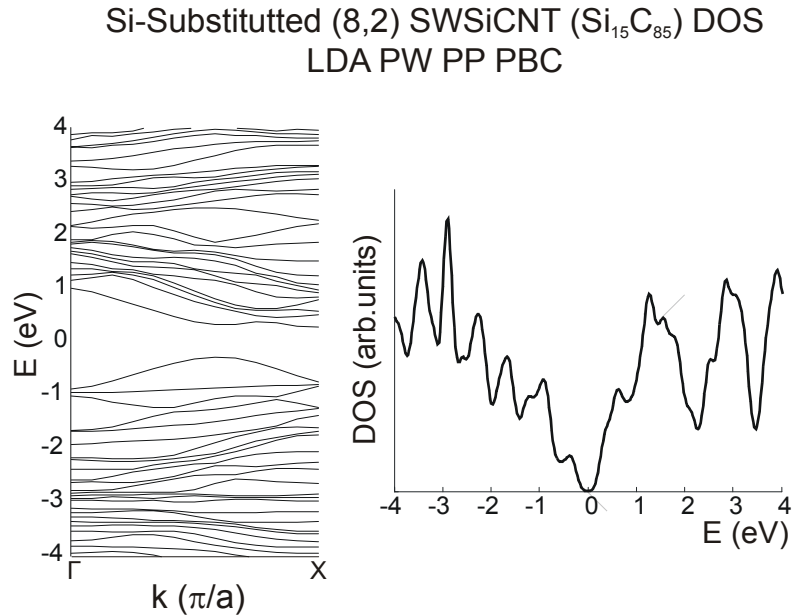

**Fig. 7. The density of the electron states and the band structure of the (8,2) SWSi$_{0.15}$C$_{0.85}$NT system.**

*c) Optical properties of Si-substituted SWCNTs*

To study the influence of silicon substitution on the quantum efficiency of the partially Si-substituted SWSi$_x$C$_{(1-x)}$NTs, we performed the TD PBE0/6-31+G** 

calculations (in the cluster approximation) of the total probabilities of the optical transitions of the fragment of the pristine and partially 11% Si-substituted of the (9,0) SWCNT as well as free carbon and silicon atoms to evaluate the role of chemical bonding and atomic factors. The integral probabilities of optical transitions (the sum over all absolute oscillator strengths) calculated using TD DFT approach for the first 12 nonzero dipole transitions for the free C and Si atoms (the excitation energies from 5.943 up to 17.141 eV for carbon and from 4.570 up to 13.041 eV for silicon) are equal to 0.807 and 1.185 a.u. correspondingly (the ratio is equal to 1.5). The main reason for such an increase in the integral probabilities is a big value of $\langle Si3p|r|Si3s \rangle$ radial dipole integrals in comparison with the $\langle C2p|r|C2s \rangle$ ones. Also, a small additional contribution in the enhancement of probability for the silicon atom is due to the presence of Si3$d$ vacant states with relatively low energies.

The same contributions over the first 24 excited states of the cluster $C_{54}H_{18}$ which is a fragment of the (9,0) SWCNT (energies of exited states from 0.1195 up to 2.5844 eV, 10 non-zero transitions) and the $Si_6C_{48}H_{18}$ (a fragment of 11% Si-substituted (9,0) SWCNT, excitation energies from 0.1548 up to 2.0384 eV, 19 nonzero transitions) show a much higher enhancement (15.7 times) of the absolute oscillator strength of the optical transitions from 0.0168 a.u ($C_{54}H_{18}$ cluster) up to 0.263 a.u. ($Si_6C_{48}H_{18}$ cluster). The PBE0/6-31+G** Mulliken charges and populations clearly show the important role of chemical bonding in enhancing the optical properties of SWSi$_x$C$_{1-x}$NT.

An averaged Mulliken configuration of carbon atoms in the pure hydrocarbon cluster is $2s^{1.1}2p^{2.8}3d^{0.1}$ (Mulliken charge 0.0), whereas the corresponding one for the silicon atoms is $2s^{1.0}2p^{2.1}3d^{0.5}$ (Mulliken charge +0.4). It should be noted that additional silicon vacant states are mainly located near the Fermi level. Summarizing, the increased population of the $d$-states in the occupied and virtual electronic states, the effects of chemical bonding and large values of $\langle Si3p|r|Si3s \rangle$ matrix elements lead to a significant increase (approximately one order of magnitude) of the absolute oscillator strength of dipole electronic transitions in the near infrared region of the Si-substituted species.

The reason of the increase of the dipole matrix elements (C $vs$ Si) is the additional core electrons in the former, effecting the increase of the spatial extension of the valence $s$ and $p$ orbitals, as well as in the greater availability of 3$d$-orbitals. The same reasons are basically responsible for the change in the band gap for the Si-substituted nanotubes.

Summarizing, the specific conditions of the high temperature synthesis of carbon nanotubes [74] lead to the generation of 1/3 of the metallic species in the carbon nanotube soot. Even a rough estimate of the enhancement of absolute probability of dipole optical transitions upon Si substitution (15%) based on atomic intensities gives a 60% enhancement of this property. The TD PBE0/6-31+G** calculations for the molecular clusters $C_{54}H_{18}$ and $Si_6C_{48}H_{18}$ predict a much larger efficiency (up to 23 times).

## V. CONCLUSIONS

Using the band structure LDA PW PP PBC calculations of the set of pristine and partially Si-substituted SWCNTs as well as the pristine SWSiCNTs, it was shown that even small silicon concentrations can greatly affect the electronic structure and band gap nature of the single wall carbon nanotubes. For the metallic armchair (8,8) SWCNT the relatively small silicon concentrations (from 5.2% of silicon) lead to an opening of the direct band gap at Γ-Γ point of Brillouin zone, whereas such concentrations for the semiconducting zigzag (10,0) SWCNT lead to a sufficient decrease of the band gap conserving its direct Γ-Γ point nature. In the range of 12 – 18% of silicon the energy of band gaps of both systems become practically the same (less than 0.5 eV).

The calculations of the (8,2) $SWSi_{0.15}C_{0.85}NT$ as an example of the chiral systems confirmed the direct band gap nature and width (0.6 eV) of the 15% Si-substituted SWCNTs. The TD DFT calculations of the absolute oscillator strength demonstrated a large (one order of magnitude) increase in the optical dipole transition probabilities in the near infrared region for the Si-containing species because of the large value of the $\langle Si3p|r|Si3s \rangle$ optical dipole matrix elements and participation of $Si3d$ states in chemical bonding in both valence and conductance bands. The combination of all properties (100% of semiconducting species, uniform direct nature and width (~0.5 eV) of the band gap and enhanced probability of electron transitions) makes the $SWSi_{0.15}C_{0.85}NTs$ promising candidates for future technological applications.

## ACKNOWLEDGMENTS

This work was supported by project "Materials Design with New Functions Employing Energetic Beams" and JAEA Research fellowship (PVA). PVA also acknowledges the personnel of Research Group for Atomic-scale Control for Novel

Materials under Extreme Conditions for hospitality and fruitful discussions. Authors acknowledge Institute of Computational Modeling SB RAS for the computer resources.

REFERENCES

[1] M. Ouyang, J.-L. Huang, C.M. Lieber, Phys. Rev. Lett. **88**, 066804-1 (2002).

[2] M. Ouyang, J.-L. Huang, C.L. Cheung, C.M. Lieber, Science **291**, 97 (2001).

[3] M. Ouyang, J.-L. Huang, C.M. Lieber, Ann. Rev. Phys. Chem. **53**, 201 (2002).

[4] P.G. Collins, K. Bradley, M. Ishigami, A. Zettl, Science **287** 1801 (2000).

[5] J. Kong, N.R. Franklin, C. Zhou, M.G. Chapline, S. Peng, K. Cho, H. Dai, Science **287**, 622 (2000).

[6] S.H. Jhi, S.G. Louie, M.L. Cohen, Phys. Rev. Lett. **85**, 1710 (2000).

[7] V.H. Crespi, M.L. Cohen, A. Rubio, Phys. Rev. Lett. **79**, 2093 (1997).

[8] J. W. G. Wildöer, L. C. Venema, A. G. Rinzler, R. E. Smalley, C. Dekker, Nature **391**, 59 (1998).

[9] P.V. Avramov, B.I. Yakobson, G.E. Scuseria, Phys. Sol. State **46**, 1168 (2004).

[10] O. Gülseren, T. Yildirim, S. Ciraci, Phys. Rev. **B66**, 121401(R) (2002).

[11] A.H. Nevidomskyy, G. Csányi, M.C. Payne, Phys. Rev. Lett. **91**, 105502-1 (2003).

[12] K.S. Kim, D.J. Bae, J.R. Kim, K.A. Park, S.C. Lim, J.-J. Kim, W.B. Choi, C.Y. Park, Y.H. Lee, Advanced Materials **14**, 1818 (2002).

[13] M. J. O'Connell, S.M. Bachilo, C.B. Huffman, V.C. Moore, M.S. Strano, E.H. Horoz, K.L. Rialon, P.J. Boul, W.H. Noon, C. Kittrell, J. Ma, R.H. Hauge, R.B. Weisman, R.E. Smalley, Science **297**, 593 (2002).

[14] F. Wang, G. Dukovic, L. E. Brus, T. F. Heinz, Phys. Rev. Lett. **92**, 177401 (2004).

[15] A. Hagen, M. Steiner, M.B. Raschke, C. Lienau, T. Hertel, H. Qian, A. J. Meixner, A. Hartschuh, Phys. Rev. Lett. **95**, 197401-1 (2005).

[16] V. Barone, J. Heyd, G.E. Scuseria, J. Chem. Phys. **120**, 7169 (2004).

[17] M.A. Lenhard, D.J. Larkin, http://www.grc.nasa.gov/WWW/RT2002/5000/5510lienhard.html

[18] X.-H. Sun, C.-P. Li, W.-K. Wong, N.B. Wong, C.-S. Lee, S.-T. Lee, B.K. Teo, J. Amer. Chem. Soc. **124**, 14464 (2002).

[19] E. Muñoz, A.B. Dalton, S. Collins, A.A. Zakhidov, R.H. Baughman, W.L. Zhou, J. He, C.J. O'Connor, B.McCarthy, W.J. Blau, Chem. Phys. Lett. **359**, 397 (2002).

[20] J.W. Liu, D.Y. Zhong, F.Q. Xie, M. Sun, E.G. Wang, W.X. Liu, Chem. Phys. Lett. **348**, 357 (2001).

[21] W. Han, S. Fan, Q. Li, W. Liang, B. Gu, D. Yu, Chem. Phys. Lett. **265**, 374 (1997).

[22] C. Pham-Huu, N. Keller, G. Ehret, M. J. Ledoux, J. Catalysis **200**, 400 (2001).

[23] T. Taguchi, N. Igawa, H. Yamamoto, S. Jitsukawa, J. Am. Ceram. Soc. **88**, 459 (2005).

[24] W. Shi, Y. Zheng, H. Peng, N. Wang, C. S. Lee, S.-T. Lee, J. Am. Ceram. Soc. **83,** 3228 (2000).

[25] Y. Zhang, T. Ichihashi, E. Landree, F. Nihey, S. Iijima, Science **285**, 1719 (1999).

[26] V.G. Sevastyanov, A.V. Antipov, V.I. Perepechenykh, B.I. Petrov, G.A. Domrachev, A.M. Ob'edkov, B.S. Kaverin, A.A. Zaitsev, K.B. Zhogova, NANOTUBE'05 Conference, Abstracts, http:/www.fy.chalmers.se/conferences/nt05/abstracts/P178.html (2005).

[27] J.-M. Nhut, R. Vieira, L. Pesant, J.-P. Tessonnier, N. Keller, G. Ehret, C. Pham-Huu, M. J. Ledoux, Catalysis Today **76**, 11 (2002).

[28] N. Keller, C. Pham-Huu, G. Ehret, V. Keller, M. J. Ledoux, Carbon **41**, 2131 (2003).

[29] E. Borowiak-Palen, M. H. Ruemmeli, T. Gemming, M. Knupfer, K. Biedermann, A. Leonhardt, T. Pichler, J. Appl. Phys. **97**, 056102 (2005).

[30] N. Keller, R. Vieira, J.-M. Nhut, C. Pham-Huu, M. J. Ledoux, J. Braz. Chem. Soc. **16**, 514 (2005).

[31] M. Lin, K. P. Loh, C. Boothroyd, A. Du, Appl. Phys. Lett. **85**, 5388 (2004).

[32] A. Mavrandonakis, G.E. Froudakis, M. Schenell, M. Mühlhäuser, Nanolett., **3**, 1481 (2003).

[33] M. Menon, E. Richter, A. Mavrandonakis, G. Froudakis, A. N. Andriotis, Phys. Rev. **B69**, 115322 (2004).

[34] M. Zhao, Y. Xia, F. Li, R.Q. Zhang, S.T. Lee, Phys. Rev. **B71**, 085312 (2005).

[35] N. Hamada, S.-i. Sawada, A. Oshiyama, Phys. Rev. Lett. **68**, 1579 (1992).

[36] R. Saito, M. Fujita, G. Dresselhaus, M.S. Dresselhaus, Phys. Rev. **B46**, 1804 (1992).

[37] S. Reich, C. Thomsen, J. Maultzsch, Carbon Nanotubes: Basic Concepts and Physical Properties, Wiley-VCH Verlag GmbH&Co. KGaA, 2004.

[38] M. Zhao, Y. Xia, R.Q. Zhang, S.-T. Lee, J. Chem. Phys. **122**, 214707 (2005).

[39] F. Li, Y.-Y. Xia, M.-W. Zhao, X.-D. Liu, B.-D. Huang, Z.-H. Yang, Y.-J. Ji, C. Song, J. Appl. Phys. **97**, 104311 (2005).

[40] R.J. Baierle, S.B. Fagan, R. Mota, A.J.R. da Silva, A. Fazzio, Phys. Rev. **B64,** 085413 (2001).

[41] A. Yakoubi, B. Bouhafs, M. Ferhat, P. Ruterana, Mat. Sci. Eng. B, **122**, 145 (2005).

[42] W. Windl, O.F. Sankey, J. Menédez, Phys. Rev. **B57**, 2431 (1998).

[43] C.-J. Lee, A. Mizel, U. Banin, M.L. Cohen, A.P. Alivisatos, J. Chem. Phys. **113**, 2016 (2000).

[44] P. Michler, A. Hangleiter, A. Moritz, G. Fuchs, V. Härle, F. Scholz, Phys. Rev. **B48**, 11991 (1993).

[45] J. Barrau, K. Khirouni, Th. Amand, J.C. Brabant, B. Brousseau, M. Brousseau, P.H. Binh, F. Mollot, R. Planel, J. Appl. Phys. **65**, 3501 (1989).

[46] G. Kresse, J. Furthmüller, Comput. Mat. Sci. **6**, 15 (1996).

[47] G. Kresse, J. Furthmüller, Phys. Rev. **B54**, 11169 (1996).

[48] P. Hohenberg, W. Kohn, Phys. Rev. **136**, B*864* (1964).

[49] W. Kohn, L. J. Sham, Phys. Rev. **140**, A*1133* (1965).

[50] D. M. Ceperley, B. J. Alder, Phys. Rev. Lett. **45**, *566* (1980).

[51] D. Vanderbilt, Phys. Rev. **B41**, *7892* (1990).

[52] A.S.Fedorov, P.B.Sorokin, Bulletin of Krasnoyarsk State University **1**, 55 (2004).

[53] A.S.Fedorov, P.B.Sorokin, Phys. Sol. State **47**, 2196 (2005).

[54] R. E. Stratmann, G. E. Scuseria, and M. J. Frisch, *J. Chem. Phys.* **109**, 8218 (1998).

[55] M. Ernzerhof, G.E. Scuseria, J. Chem. Phys. **110**, 5029 (1999).

[56] J.P. Perdew, M. Ernzerhof, K. Burke, J. Chem. Phys. **105**, 9982 (1996)

[57] C. Adamo, V. Barone, J. Chem. Phys. **110**, 6158 (1999).

[58] Gaussian 03, Revision C.02, M. J. Frisch, G. W. Trucks, H. B. Schlegel, G. E. Scuseria, M. A. Robb, J. R. Cheeseman, J. A. Montgomery, Jr., T. Vreven, K. N. Kudin, J. C. Burant, J. M. Millam, S. S. Iyengar, J. Tomasi, V. Barone, B. Mennucci, M. Cossi, G. Scalmani, N. Rega, G. A. Petersson, H. Nakatsuji, M. Hada, M. Ehara, K. Toyota, R. Fukuda, J. Hasegawa, M. Ishida, T. Nakajima, Y. Honda, O. Kitao, H. Nakai, M. Klene, X. Li, J. E. Knox, H. P. Hratchian, J. B. Cross, V. Bakken, C. Adamo, J. Jaramillo, R. Gomperts, R. E. Stratmann, O. Yazyev, A. J. Austin, R. Cammi, C. Pomelli, J. W. Ochterski, P. Y. Ayala, K. Morokuma, G. A. Voth, P. Salvador, J. J. Dannenberg, V. G. Zakrzewski, S. Dapprich, A. D. Daniels, M. C. Strain, O. Farkas, D. K. Malick, A. D. Rabuck, K. Raghavachari, J. B. Foresman, J. V. Ortiz, Q. Cui, A. G. Baboul, S. Clifford, J. Cioslowski, B. B. Stefanov, G. Liu, A. Liashenko, P. Piskorz, I. Komaromi, R. L. Martin, D. J. Fox, T. Keith, M. A. Al-Laham, C. Y. Peng, A. Nanayakkara, M. Challacombe, P. M. W. Gill, B. Johnson, W. Chen, M. W. Wong, C. Gonzalez, and J. A. Pople, Gaussian, Inc., Wallingford CT, 2004.

[59] P.V. Avramov, K.N. Kudin, G.E. Scuseria, Chem. Phys. Lett. **370**, 597 (2003).

[60] M. Machón, S. Reich, C. Thomsen, D. Sánchez-Portal, P. Ordejón, Phys. Rev. B **66**, 155410 (2002).

[61] R. Bauernschmitt, R. Ahlrichs, Chem. Phys. Lett. **256**, 454 (1996).

[62] Z. Zhou, M. Steigerwald, M. Hybertsen, L. Brus, R.A. Friesner, J. Amer. Chem. Soc. **126**, 3597 (2004).

[63] Z.M. Li, Z.K. Tang, H.J. Liu, N. Wang, C.T. Chan, R. Saito, S. Okada, G.D. Li, J.S. Chen, N. Nagasawa, S. Tsuda, Phys. Rev. Lett. **87**, 127401 (2001).

[64] H.J. Liu, C.T. Chan, Phys. Rev. B **66**, 115416 (2002).

[65] J.W. Minitmire, C.T. White, Synth. Met. **77**, 231 (1996).

[66] C.D. Spataru, S. Ismail-Beigi, L.X. Benedict, S.G. Louie, Phys. Rev. Lett **92**, 077402 (2004).

[67] G. Onida, L. Reining, A. Rubio, Rev. Mod. Phys. **74**, 601 (2002).

[68] A.G. Marinopoulos, L. Reining, A. Rubio, N. Vast, Phys. Rev. Lett. **91**, 046402 (2003).

[69] J. W. Mintmire, B. I. Dunlap, C. T. White, Phys. Rev. Lett. **6**8, 631 (1992).

[70] P. Käckell, B. Wenzien, F. Bechstedt, Phys. Rev. B **50**, 10761 (1994).

[71] W.R.L. Lambrecht, in Diamond, SiC and Nitride Wide Bandgap Semiconductors, edited by C.H. Carter, Jr., G. Gildenblat, S. Nakamura, and R.J. Nemanich, MRS Symposia Proceedings No 339 (Materials Research Society, Pittsburg, 1994), p. 565.

[72] L. Wenchang, Z. Kaining, X. Xide, J. Phys. Condens. Matter **5**, 883 (1993).

[73] B. Wenzien, P. Käckell, F. Bechstedt, Phys. Rev. B **52**, 10897 (1995).

[74] A.V. Eletskii, Physics-Uspekhi **40**, 899 (1997).